\documentclass[11pt,a4paper]{article}
\usepackage{epsfig}

\newlength{\dinwidth}
\newlength{\dinmargin}
\def\eq#1{{(\ref{#1})}}
\newcommand{\Le}{\left(}
\newcommand{\Ra}{\right)}

\newcommand{\bb}{\mathbf{b}}

\newcommand{\beq}{\begin{equation}}
\newcommand{\eeq}{\end{equation}}
\newcommand{\beqar}[1]{\begin{eqnarray}\label{#1}}
\newcommand{\eeqar}{\end{eqnarray}}
\newcommand{\tdm}[1]{\mbox{\boldmath $#1$}}
\newcommand{\tdmi}[1]{\mbox{\boldmath\scriptsize $#1$}}

\newcommand{\vb}{\tdm{b}}
\newcommand{\vq}{\tdm{q}}
\newcommand{\vk}{\tdm{k}}

\newcommand{\vbb}{\tdmi{b}}
\newcommand{\vqq}{\tdmi{q}}

\newcommand{\im}{\mathrm{Im}\;}
\begin{document}

\title {
{\Large \bf Exclusive Higgs boson production at the LHC:\\
hard rescattering corrections}}
\author{ 
{~}\\
{~}\\
{\large 
J.~Bartels$\,{}^{a)}\,$\thanks{Email: bartels@mail.desy.de},
\hspace{1ex}
S.~Bondarenko$\,{}^{a)}\,$\thanks{Email: sergb@mail.desy.de}, 
\hspace{1ex}
 K.~Kutak $\,{}^{a),b)}\,$\thanks{E-mail: kutak@mail.desy.de} 
\hspace{0.5ex} 
and 
\hspace{0.5ex}
 L.~Motyka$\,{}^{c),d)}\,$\thanks{E-mail: motyka@th.if.uj.edu.pl}}
 \\[10mm]
{\it\normalsize ${}^{a)}$ II Institute for  Theoretical Physics, 
   University of Hamburg, Germany}\\
{\it\normalsize ${}^{b)}$ H.~Niewodnicza\'{n}ski Institute of Nuclear Physics,
 Krak\'{o}w, Poland}\\
{\it\normalsize ${}^{c)}$ DESY Theory Group, Hamburg, Germany }\\
{\it\normalsize $^{d)}$ Institute of Physics, Jagellonian University,
Krak\'{o}w, Poland} \\}

\date{January 16, 2006} 

\maketitle
\thispagestyle{empty}

\begin{abstract}
We examine, as a correction to the central exclusive Higgs boson production 
in $\,pp\,$ collisions at the LHC, the rescattering of gluonic ladders off 
the proton. As usual, at the lowest order the hard
part of this process can be described as a fusion of two hard gluonic ladders. 
We calculate corrections to this hard amplitude which are due to 
rescattering of these ladders. 
These corrections, which contain high mass diffractive 
excitations of the proton, have not yet been taken into 
account by the usual soft survival probabilities.
We find that the correction due to the exchange of a single hard 
rescattering is negative, large and infrared sensitive. 
As a first step towards a more reliable description we 
therefore replace the rescattering exchange by a unitarized amplitude 
using the BK equation which generates the saturation scale $Q_s(x)$.
We also include a soft gap survival probability factor. 
We discuss the results and outline possible future strategies.
\end{abstract}

\begin{flushright}
\vspace{-20.8cm}
{DESY 06--003}\\
hep-ph/0601128  \\
\end{flushright}
\thispagestyle{empty}

\newpage

\section{Introduction}

One of the main experimental goals to be achieved at the Large Hadron
Collider~(LHC)  is the discovery of the Higgs boson and the  investigation of
its properties. The discovery potential of LHC should be sufficient to 
find the Standard Model Higgs boson as well as the supersymmetric 
Higgs particles in the wide range of masses, starting from the lower limit of 
114.4~GeV imposed by the LEP~\cite{pdg} data up to the mass of a few 
hundred GeV. Precise electroweak measurements indicate, however, 
that the mass of the Higgs boson should be rather at the
lower end of the allowed window, with the central value at 113~GeV being 
determined from radiative corrections, and the upper limit of about 240~GeV at
the confidence level of 95\%. 
It has been proposed~\cite{SNS}--\cite{Bzdak}
that, if the Higgs boson mass, $M_H$, is close to the expected central
value, a measurement of the exclusive diffractive reaction 
should be possible:
\begin{equation}
p\; p \;\; \longrightarrow \;\; p\; H \; p,
\end{equation}
with the Higgs boson separated by rapidity gaps from the protons.
The main advantage of this process is an extremely low background which is much
smaller than the potential signal. The background suppression permits
precise measurements -- for instance of the quantum numbers of the Higgs
boson. Besides that, in the exclusive production process, detection of
protons provides a complete information on the invariant mass of the
produced system. This opens an exciting possibility to find resonances
corresponding to new heavy particles that would be invisible in conventional
measurements, see~\cite{KMR8,KMR9,KMR10}.
The main drawback of the exclusive Higgs production channel, however, is
its very low rate. The requirement of leaving the protons intact and of 
producing no secondary particles reduces the cross section by many orders of 
magnitude. In the series of papers~\cite{KMR1}-\cite{KMR7} 
the authors discussed extensively the theoretical and phenomenological 
aspects of the exclusive production of heavy particle in $\,pp\,$ 
collisions in perturbative QCD, and their approach will be used as the 
basis of our study.

\begin{figure}[t]

\begin{center}
\leavevmode
\begin{minipage}{6cm}
{\Large\bf a)}\\
\epsfxsize=6cm
\epsfysize=5cm
\epsffile{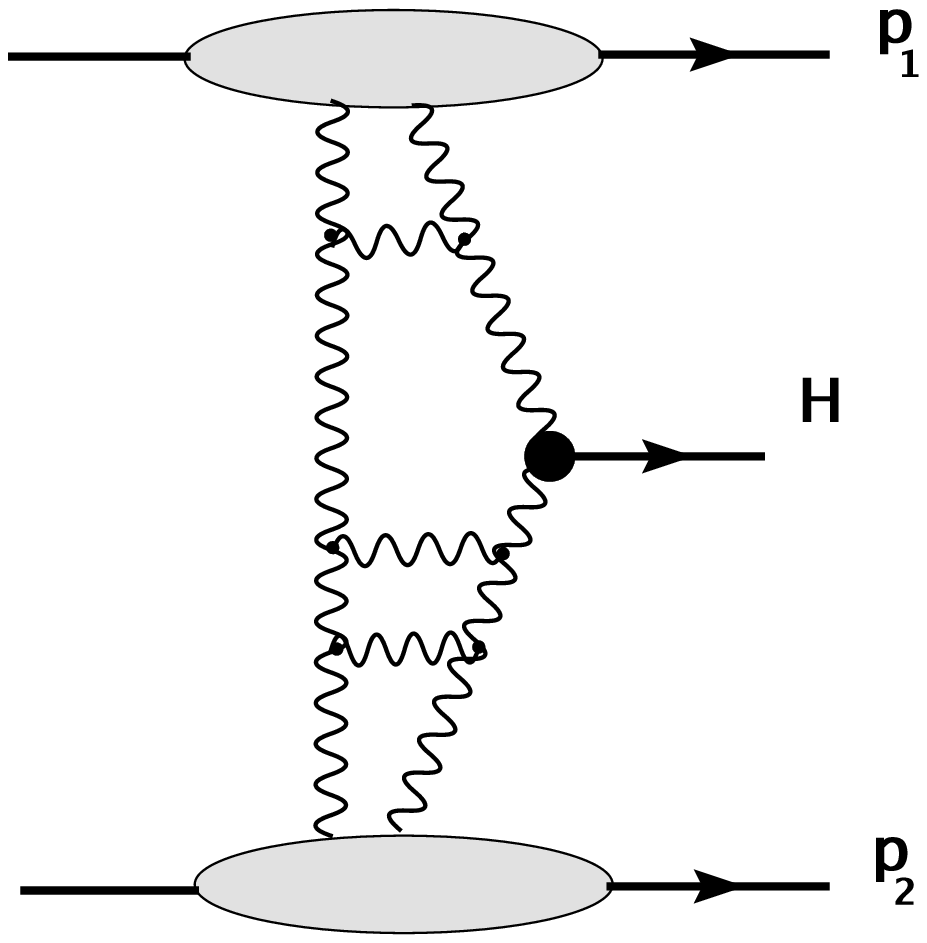}
\end{minipage}
\hspace{2cm}
\begin{minipage}{6cm}
\epsfxsize=6cm
\epsfysize=5cm
{\Large\bf b)} \\
\epsffile{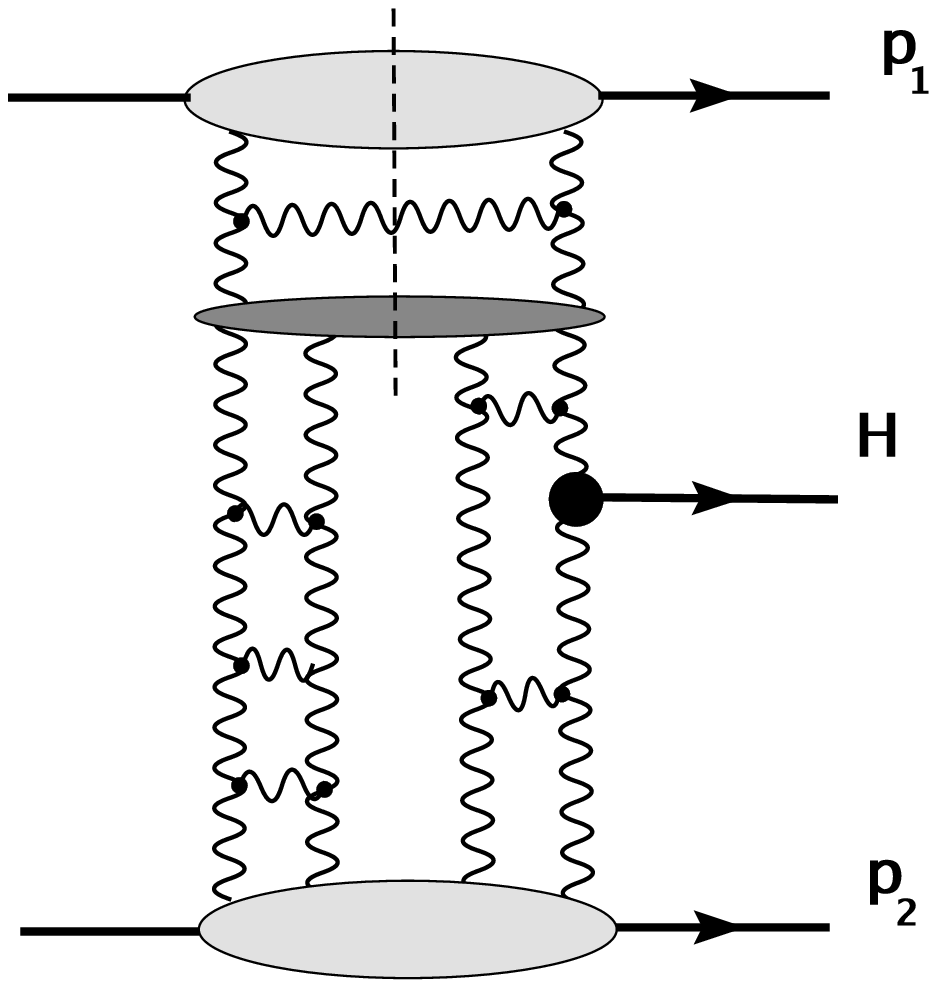}
\end{minipage}
\end{center}
\caption{\it The exclusive Higgs boson production in a $pp$  collision:
a) the Two Pomeron Fusion contribution and b) a hard absorptive correction.}
\label{Ampl0}
\end{figure}

The mechanism of the considered process is illustrated in the diagram in
in Fig.~\ref{Ampl0}a. Thus, the hard subprocess is given by a fusion of two
colour singlet gluonic ladders into the Higgs scalar. 
The diagram may be evaluated in the $k_t$-factorisation framework using 
the unintegrated gluon densities.
The required absence of hard  radiation leaves un-compensated the virtual
corrections to the process.
Those corrections turn out to be large as they contain the scale of the
Higgs boson mass generating large double and single logarithmic
contributions. The necessary resummation of these logarithmically
enhanced terms is realized by including the Sudakov form factor, which not 
only reduces the cross section but also provides the infrared stability
of the hard exclusive scattering amplitude. Unfortunately, 
the basic hard subprocess is accompanied by multiple soft rescattering,
which, typically, is inelastic and suppresses the exclusive cross section.
Currently, the soft inelastic rescattering may  be 
estimated only in phenomenological models~\cite{GLM,KMR11,KMR12,SBL}. 
Present calculations of the gap survival probability rely predominantly 
on a two channel eikonal model for the  $\,pp\,$ opacity, incorporating 
effects of the low mass diffraction and of the pionic cloud.
The model was developed in Refs.~\cite{KMR11,KMR12}, and the parameters were 
fitted to describe the data on the total $\,pp\,$ and $\,p\bar p\,$ 
cross section and on the elastic $\,pp\,$ and $\,p\bar p\,$ scattering.

In this paper we shall address the problem of computing a hard absorptive correction
to the exclusive production process, which has already been mentioned 
in the literature~\cite{KMR10} but has not been studied in detail yet. 
In brief, this contribution comes from an absorption of the hard gluonic 
ladders that fuse to produce the Higgs boson in the $\,pp\,$ scattering 
process, see Fig.~{\ref{Ampl0}b.
This contribution can also be viewed as the rescattering of a large mass 
intermediate diffractive state moving along one of the protons 
(the proton at the top of in Fig.~{\ref{Ampl0}b; let us  call it the 
{\em projectile}) off the other proton (the proton at the bottom of in 
Fig.~{\ref{Ampl0}b, the {\em target}), and this contribution is not part 
of the soft rescattering of the spectators that contributes to the standard 
opacity. In Fig.~\ref{Ampl0}b, the large mass diffractive state is indicated 
by a dashed line.
In perturbative QCD, within the leading logarithmic~$1/x$~approximation,  
the rescattering is driven by the exchange of a BFKL Pomeron 
~\cite{bfkl1,bfkl2,bfkl3,bfklsum} which couples to the hard gluonic ladder 
by the triple Pomeron vertex~\cite{JB,BW,MB,BRV}. 
Such a contribution is potentially large, since the BFKL rapidity 
evolution between the {\em target} proton and the triple Pomeron vertex
extends over the length of $Y>14$ and thus may generate an enhancement 
by two orders of magnitude. If this correction is really as large as 
anticipated, it is clear that also higher order unitarity 
corrections to this hard rescattering will be sizable and should be 
evaluated as well.

As we shall discuss in detail, the situation is even more complicated.
Namely, when evaluating the rescattering correction based on a simple BFKL 
Pomeron exchange, it turns out that the amplitude -- which is infrared finite 
-- exhibits a high sensitivity to the details of the infrared region. 
To be more precise,
the momentum scale around the triple Pomeron vertex tends to be small.   
At first sight, therefore, this na\"{i}ve approach seems to be highly 
questionable. However, if we include -- perturbative -- unitarity 
corrections, the problems with the infrared sensitivity may be greatly 
reduced. As a guiding example let as quote results of Refs.~\cite{AB,GBMS} 
on the impact of a special class of unitarity corrections on the 
behaviour of the gluon distribution at low momenta. 
In these references the Balitsky-Kovchegov (BK) equation 
\cite{Bal,Kov1,Kov2} was investigated, as a model which unitarizes the BFKL 
Pomeron exchange by a resummation of the BFKL Pomeron fan diagrams. 
It was found that the equation generates a momentum scale, the saturation 
scale $Q_s(Y)$, which provides an effective lower cut-off on the gluon momenta.
The saturation scale grows exponentially with the evolution 
length $Y$, $Q_s(Y) \sim \exp(\lambda Y)$, with $\lambda \simeq 0.3$, 
and the saturation scale enters the region in which perturbative QCD 
is applicable if~$Y$ is sufficiently large. Thus, in the presence of the
perturbative saturation scale the high sensitivity to the infrared domain
is eliminated. Let us stress here that the concept of 
a rapidity dependent saturation scale is rather robust. 
The concept has received strong support from phenomenological considerations, 
being the central part of the very fruitful saturation model~\cite{GBW1,GBW2}.
It was also proven that the generation of the saturation scale is a universal
phenomenon in a wide class of non-linear evolution 
equations~\cite{MuPe1,MuPe2,MuPe3}, which gives some confidence that 
the emergence of the saturation scale is not an accidental feature 
of the BK equation.

In the second part of our investigation, therefore, 
we shall replace the linear BFKL equation for the screening Pomeron  
by the non-linear Balitsky-Kovchegov evolution equation.
This corresponds to the resummation of the BFKL Pomeron 
fan diagrams originating from the {\em target} proton: 
this resummation will be shown to stabilize the correction amplitude 
in the infrared region. 
Certainly, in doing so we have made rather crude approximations. 
Namely, we have not taken into account other important 
unitarity corrections present in the case of proton-proton
scattering, for instance those related to possible interactions of the
Pomerons in the fan diagrams with the {\em projectile} proton (for a more
detailed discussion see Section~7). For the total cross section of 
nucleus-nucleus collisions an effective field theory of interacting 
BFKL Pomerons has been formulated ~\cite{MB2,MB3,MB4}. For $pp$ scattering
an analogous framework is still to be developed.       
We therefore view our study only as a first step towards constructing a 
theoretical framework for evaluating hard absorptive corrections 
to the exclusive high mass production in $\,pp\,$ collisions. 
We shall compute the magnitude of the correction in this, 
rather crude, approximation in order to provide some insight into 
how important the discussed effects might be.

The structure of the paper is the following. In Section~2 we
briefly review the standard framework introduced in the series
of papers~\cite{KMR1}--\cite{KMR7}. We derive the hard rescattering 
correction in Section~3. The calculation of the soft gap survival 
probability for the hard corrected amplitude is given in Section~4.  
In Section~5 the Balitsky-Kovchegov evolution of the gluon density
is described. The results of the numerical evaluation of the hard 
rescattering corrections and the obtained cross sections are 
presented in Section~6. Implications of the results and future strategies 
are discussed in Section~7. Finally, conclusions are presented in Section~8.


\section{The hard amplitude for exclusive Higgs production}

In this Section we briefly review the main features of the
standard QCD approach~\cite{KMR1}--\cite{KMR7} to the hard part of the
exclusive Higgs boson production in $\,pp\,$ collisions.

The basic diagram is given in Fig.~\ref{Ampl1}; the leading contribution 
of this amplitude is imaginary.
The diagram is evaluated in the high energy limit, where the $t$-channel 
gluons are in the so-called nonsense polarisation states, and their 
virtualities are dominated by the transverse components of their momenta.
The matrix element describing the fusion of two gluons into the Higgs
boson is given by the top quark triangle, taken in the limit of a 
point-like coupling. This effective approximation is very accurate as the gluon
virtualities are much smaller than the top quark mass.
The amplitude of finding, in the proton, two gluons that carry tranverse
momenta $\vk$ and $\vk'$ and fractions $x$ and $x'$
of the longitudinal momentum of the parent proton is represented by
the off-diagonal unintegrated gluon density
$f^{\mathrm{off}} _g (x,x';\vk,\vk';\mu)$, assuming
that the density is probed at the scale $\mu$.
The three sub-amplitudes (Fig.~\ref{Ampl1}), corresponding 
to the gluon emissions and to the Higgs
boson production, are convoluted using the $k_t$-factorisation.

\begin{figure}[t]
\begin{minipage}{11.0 cm}
\begin{center}
\epsfxsize=7cm
\epsfysize=7cm
\leavevmode
\hbox{\epsffile{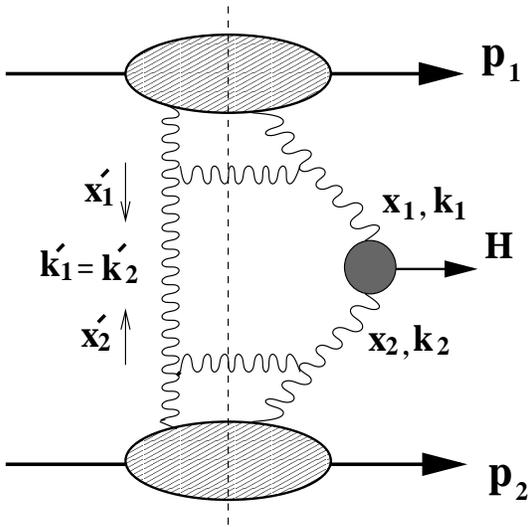}}
\end{center}
\end{minipage}
\begin{minipage}{5.0cm}
\caption{\it Kinematics of the Two Pomeron Fusion contribution to the
exclusive Higgs boson production.}
\label{Ampl1}
\end{minipage}
\end{figure}

The kinematics of the exclusive Higgs boson production determines
the arguments of the off-diagonal gluon densities. Thus, the elastic
form-factor of the proton provides a sharp cut-off, which we assume to
be Gaussian:
$\;\sim\exp[- (\vk-\vk')^2 R^2 /4]$ with $R^2 \simeq 8$~GeV$^2$,
on the momentum transfer carried by the gluonic system.
Therefore, it is appropriate to write
\beq
f^{\mathrm{off}} _g (x,x';\vk,\vk';\mu) \simeq
f^{\mathrm{off}} _g (x,x';k^2;\mu) \,
 \exp[- (\vk-\vk')^2 R^2 / 4].
\label{fqdep}
\eeq
For future use, let us also define the unintegrated gluon distribution
depending on the position~$\vb$ in the transverse plane
\beq
\label{fbdep}
\tilde f^{\mathrm{off}} _g (x,x';k^2,\vb;\mu) =
\int {d^2 q  \over (2\pi)^2}\; e^{i\vqq \vbb} \;
f^{\mathrm{off}} _g (x,x';\vk,\vq-\vk;\mu),
\eeq
where $\vq = \vk - \vk'$.

The next point that should be discussed is the disparity between the
longitudinal momentum fractions
$x$ and $x'$. The production of the heavy scalar particle occurs close to 
the central region where $x\simeq M_H /\sqrt{s}$ and $x' \sim \sqrt{k^2 / s}$, so $x' \ll x$.
At the leading logarithmic level the off-diagonal (w.r.t. $x$) distribution
reduces to the diagonal one. Beyond the LL accuracy the effect
of different $x$ values of the gluons may be approximated by a constant
factor~\cite{SHUV}
\beq
R_{\xi} = {2^{2\lambda+3} \over \sqrt\pi} {\Gamma(\lambda+5/2) \over
\Gamma(\lambda+4)},
\label{roff}
\eeq
provided the diagonal gluon distribution has a power-like $x$-dependence,
$f_g(x,k^2;\mu) \sim x^{-\lambda}$. For moderate scales one has 
$\,\lambda \simeq 0.25$.

The last important point is the scale dependence of the unintegrated gluon 
density. In the BFKL approach to the evolution of the gluon density 
there is no dependence of the gluon density on the external scale. 
This happens because the virtual perturbative QCD
corrections to the coupling of the ladder to the Higgs production vertex 
generate only the DGLAP logarithms $\log(\mu^2 / k^2)$ (double and single) 
and not the small~$x$ logarithms. In double logarithmic (in $\log (\mu^2/k^2)$)
accuracy, the large corrections may be resumed into the Sudakov
form-factor~\cite{KMR2}:
\beq
T_s(k,\mu) = \exp\biggl(
-\int_{k ^2} ^{\mu^2} {dp^2 \over p^2} { N_c \alpha_s \over \pi}
\int_{p} ^{\mu} {d\omega \over \omega}  \biggr).
\label{2lsud}
\eeq
The calculation of the Sudakov form-factor at single logarithmic accuracy
has also been performed~\cite{KMR2}, giving
\beq
T_g (\vk,\mu) =
\exp\left( -\int_{{k}^2}^{\mu^2} \frac{dp^2}{p^2}
\frac{\alpha_s(p^2)}{2\pi}
\int_0^{1-\delta} dz z \left[P_{gg}(z)+\sum_q P_{qg}(z)\right]\right) ,
\label{sudakov}
\eeq
where $\delta = \frac{p}{p+\mu}$ is chosen to provide the correct angular
ordering of the real gluon emissions, and $P_{gg}$ and $P_{qg}$ are the
DGLAP leading order splitting functions.
For the exclusive Higgs production the scale $\mu \simeq 0.62 M_H$ should
be chosen~\cite{KMR7} in order to match exactly the first order single
logarithmic correction. In our calculation we shall adopt the scale 
$\mu = M_H/2$, which does not introduce a significant change 
of the result compared to the choice of~\cite{KMR7}.

A thorough discussion of how to determine the two-scale diagonal
unintegrated gluon distribution has been presented in 
Refs.~\cite{KimMR1,KimMR2}.
A simple and economic way to incorporate the Sudakov form-factor into the
unintegrated gluon distribution is given by the approach in which the last
step of the DGLAP evolution is resolved in terms of the transverse momentum.
In this approach, the two-scale unintegrated gluon distribution is 
obtained from the collinear gluon distribution in the following way~\cite{DDT,KimMR1,KimMR2}:
\beq
f_g(x,k^2;\mu) =Q^2\frac{\partial}{\partial Q^2}
\biggl[x g(x,Q^2)\cdot T_g(Q,\mu)\biggr]_{Q^2=k^2}.
\label{unglue}
\eeq
In the off-diagonal case which is relevant for our present study, 
the logarithmic virtual corrections
are associated with the production vertex at which the pair of gluons 
merges (see Fig.~\ref{Ampl1}), in contrast to the standard case
of a diagonal gluon distribution in DIS or in a hard inclusive production 
processes. Thus, in \eq{unglue} the Sudakov form-factor has to be replaced 
by its square root. 

Collecting all these results we arrive at:
\beq
\label{kmrglue}
f_g ^\mathrm{off} (x,k^2;\mu) =
R_{\xi}  Q^2 \frac{\partial}{\partial Q^2} \;
\biggl[ x g(x,Q^2)\cdot \sqrt{ T_g(Q,\mu)} \biggr]_{Q^2=k^2}.
\eeq
After contracting the polarisation tensors of the gluons
with the $ggH$ vertex we obtain the dominant imaginary part of the 
hard exclusive amplitude for $\; pp \,\to \, p\, H \, p\,$ 
in the forward direction in the following form:
\beq\label{born}
\im M_{0}(y) = 2\pi^3 A \;
\int\,\frac{d^2k}{2\pi k^4}\,
f_g ^{\mathrm{off}}(x_1,k^{2};\mu)\,
f_g ^{\mathrm{off}}(x_2,k^{2};\mu).
\eeq
Here the constant $A$ is given by
\beq\label{koeff}
A^2\,=K\,\sqrt{2}\,G_F\,\alpha_{S}^{2}(M_{H}^{2})\,/9\,\pi^2\, ,
\eeq
where $\,G_F\,$ is the Fermi coupling, and $\,K\approx\,1.5\,$
is the NLO $K$-factor.
The Higgs boson rapidity in the centre of mass frame, $y$, defines the
values of longitudinal momentum fractions:
\beq
x_{1,2} = {M_H \over \sqrt{s}} \, \exp(\pm y).
\eeq
The differential hard cross-section following from this amplitude
takes the form:
\beq
\left.
{d\sigma_{pp\to pHp}(y) \over dy\, dt_1\, dt_2}\right|_{t_1=t_2=0}  =
{|M_0(y)|^2 \over 256\pi^3}.
\eeq
After the inclusion of the proton elastic form-factors in the non-forward
direction (parameterized by $\;\sim\exp[- (\vk-\vk')^2 R^2 /4]$ in
\eq{fqdep}) one obtains the dependence of the cross section
on $t_1$ and $t_2$ of the form of $\exp(R^2(t_1+t_2)/2)$. 
Thus, after the integration over momentum transfers $t_1$ and $t_2$, 
the single differential cross section reads:
\beq
{d\sigma_{pp\to pHp}(y) \over dy} =  {|M_0(y)|^2 \over 64\pi^3 R^4}.
\eeq
Note that, in (\ref{born}), the form of the integral over the transverse 
momentum suggests an important sensitivity to the infrared domain.
The rescue comes from the Sudakov form-factor which suppresses the
low momenta of the unintegrated gluons faster than any power of $k$.
It turns out that the dominant contribution to the integral
(\ref{born}) comes from momenta $k > 1$~GeV.

More details about the derivation of the amplitude of the exclusive
Higgs boson production are given in the series of  
papers~\cite{KMR1}--\cite{KMR7}. 
In what follows we shall refer to the amplitude 
(\ref{born}) as to the Two Pomeron Fusion~(TPF) Amplitude.


\section{Hard rescattering corrections}
\subsection{Construction of the amplitude}

The amplitude (\ref{born}) includes resummed QCD corrections up to 
single logarithms of the Higgs boson mass. The large available rapidity
gap between the scattered protons ($Y_{\mathrm{tot}} \sim 19$ for the LHC 
energies), however, suggests that significant QCD
corrections enhanced by small~$x$ logarithms should be also considered.
The spectator rescattering is expected to be of non-perturbative nature,
and commonly it is accounted for by a soft gap survival probability factor.

Fig.~\ref{Ampl2} illustrates a process of
absorption of the perturbative gluonic ladder from the {\em projectile} 
(upper) proton in the {\em target} (lower) proton by the exchange of a  
QCD Pomeron across a large rapidity gap.
Let us stress that the evolution length in rapidity of the screening 
Pomeron is large only if it couples to the ladder between the {\em projectile} 
proton and the produced Higgs boson. 
To be more specific, for the central Higgs boson production at the LHC
the rapidity distance between the hard ladder that originates from
one proton, and the other proton, $y_4\,=\,\log (1/x_4) > 14$, is large
while the rapidity range occupied by each of the two hard ladders, 
is much smaller, $y_1\,=\,\log (1/x_1)  < 5$ and 
$y_3\,=\,\log (1/x_3)  < 5$.
In contrast to this, the rapidity distance $y_4$ can be much larger, 
and a small~$x$ (BFKL~\cite{bfklsum}) resummation of the  perturbative QCD 
series may lead to a large enhancement
of the amplitude $\;\sim \exp(\lambda y_4)$ with $\lambda \simeq 0.3$.
Therefore this correction may be sizable, despite the fact it is suppressed
by the factor of $\alpha_s ^2$ at the triple Pomeron vertex,
illustrated in Fig.~\ref{TPV}. 
Thus, we shall evaluate the absorptive correction
corresponding to the rescattering of the hard {\em projectile} gluonic
system off the {\em target} proton. The coupling of the screening Pomeron 
to the screened gluonic ladder is known in the high energy limit; 
it is derived from the triple Pomeron vertex~\cite{JB,BW,MB,BRV}.

\begin{figure}[t]
\begin{minipage}{8.5 cm}
\begin{center}
\epsfxsize=6.5 cm
\leavevmode
\hbox{\epsffile{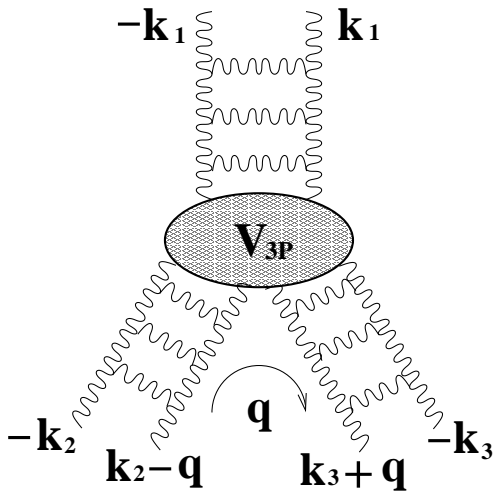}}
\end{center}
\end{minipage}
\begin{minipage}{6.5cm}
\caption{\it The triple Pomeron vertex.}
\label{TPV}
\end{minipage}
\end{figure}

The triple Pomeron vertex in pQCD has been computed in the 
LL$1/x$~approximation~\cite{JB,BW}, and it follows from the $2 \to 4$ gluon 
vertex by projecting on the colour singlet
states of the diagram represented symbolically in Fig.~\ref{TPV}.
In the process of interest the momentum transfers of all three 
Pomerons are small: for the upper Pomeron and for the lower left one 
they are damped by the elastic proton form-factors. We, therefore, will 
neglect them in the expression for the triple Pomeron vertex, and it is 
will be sufficient to evaluate the vertex in the triple-forward limit.
In this limit, the convolution $[V_{3P} \otimes D]$
of the triple Pomeron vertex $V_{3P}$ with the two-gluon amplitude 
above the vertex (here we follow the notation of~\cite{BW}, below we will 
replace $D$ by the unintegrated gluon density) is given by~\cite{JB,BW}:
\begin{eqnarray}\label{Triple}
& &
{[V_{3P} \otimes D](\vk_2,\,\vk_3)\,= \,}
\nonumber \\
& &
\alpha_{s}^{2}\, c_{3P}\;
\int {d^2 \vk_1 \over 2\pi k_1^2}\;
\Biggl\{\,
\Le\,\frac{k_{2}^{2}\,}{(\vk_{1}-\vk_{2})^2}+
\frac{k_{3}^{2}\,}{(\vk_{1}+\vk_{3})^2}-
\frac{(\vk_{2}+\vk_{3})^2\,k_{1}^{2}}
{(\vk_{1}-\vk_{2})^2\,(\vk_{1}+\vk_{3})^2}\,
\Ra\,D(k_1)\,
\nonumber \\
& & \,-\,\,\frac{k_{1}^{2}}{(\vk_{1}-\vk_{2})^2}\Le\,
\frac{k_{2}^{2}}{(\vk_{1}-\vk_{2})^2+k_{1}^{2}}-
\frac{(\vk_{2}+\vk_{3})^2}
{(\vk_{1}-\vk_{2})^2\,+\,(\vk_{1}+\vk_{3})^2}\,
\Ra\,D(k_2)\,
\nonumber \\
& & \,-\,\,\frac{k_{1}^{2}}{(\vk_{1}+\vk_{3})^2}\Le\,
\frac{k_{3}^{2}}{(\vk_{1}+\vk_{3})^2+k_{1}^{2}}-
\frac{(\vk_{2}+\vk_{3})^2}
{(\vk_{1}-\vk_{2})^2\,+\,(\vk_{1}+\vk_{3})^2}\,
\Ra\,D(k_3)\,\Biggr\}\, ,
\end{eqnarray}
where
$\pm\vk_2$ and $\pm\vk_3$ are gluon momenta
inside the two Pomerons below the vertex (see Fig.~\ref{TPV}).
Note that, in formula (\ref{Triple}), the triple Pomeron vertex 
includes both real (the first term in the l.h.s.) 
and virtual (the next term in the l.h.s.) 
contributions. The normalisation factor, $c_{3P}$, of the vertex
will fixed by the form of the final expression for the 
correction amplitude (\ref{corr}), and we have constructed this expression 
in such a way that $\; c_{3P}=1$.

\begin{figure}[t]
\begin{minipage}{12.0 cm}
\begin{center}
\epsfxsize=10cm
\leavevmode
\hbox{ \epsffile{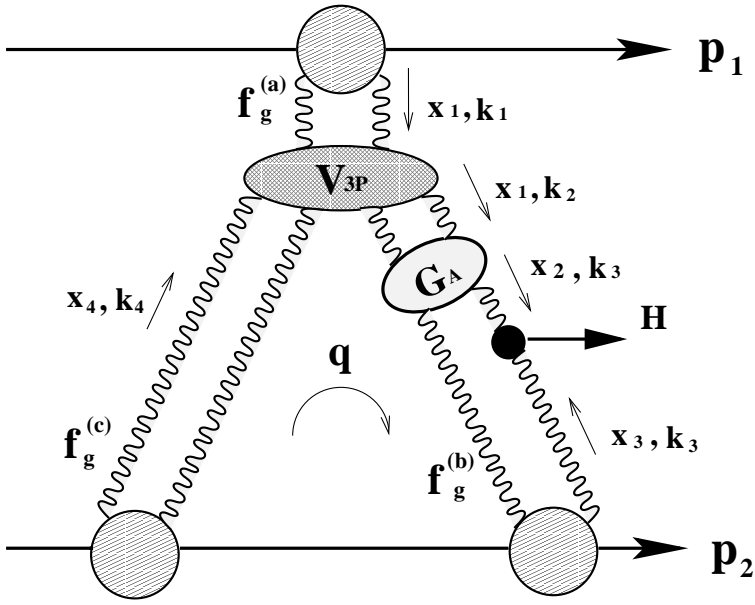}}
\end{center}
\end{minipage}
\caption{\it The hard rescattering correction to the exclusive
Higgs boson production.}
\label{Ampl2}
\end{figure}

In order to build up the complete expression of the amplitude of the
diagram shown in Fig.~\ref{Ampl2} it is necessary to take into account the
following building blocks:

\begin{enumerate}

\item The unintegrated gluon distribution, $f^{(a)}_g(x_1,k_1 ^2;\mu_a)$,
describing the amplitude of the gluons above the triple Pomeron vertex 
$V_{3P}$. This gluon distribution is evaluated at relatively large $x_1$, 
and the momentum scale in the vertex should be given by the typical 
virtualities of the vertex, $\mu_a \simeq k_1$.
So the gluon distribution will be evaluated using
\eq{unglue}\footnote{Note that, for $\mu_a = k_1$, the Sudakov
form factor, $T_g(k_1,\mu_a)$, in (\ref{unglue}) introduces only a minor 
effect.}, and we shall denote the convolution (\ref{Triple})
of the triple Pomeron vertex with the unintegrated gluon,
$f^{(a)}_g(x_1,k_1;k_1)$, by
\beq
[V_{3P} \otimes f^{(a)} _g (x_1)](k_2,k_3). 
\eeq

\item The two-gluon Green's function $G_A(\vk_2,\vk_3;x_1,x_2;\mu_a,\mu_b)$
that describes the propagation of the two gluon singlet between the
Higgs production vertex and the triple Pomeron vertex. Two
approximations for the Green's functions will be considered:
the two gluon approximation ($G_{0} (\ldots)$) and the diffusion
approximation to the BFKL evolution ($G_{\mathrm{BFKL}} (\ldots)$).
The Green's function will have the typical scales
$\mu_a$ and $\mu_b$, where the latter scale belongs to the Higgs production 
vertex. As to the scale of the triple Pomeron 
vertex, $\mu_a \sim k_2$, a priori, we do not know whether it will be large.
However, the large scale is certainly present in the Higgs vertex, 
$\mu_b \simeq M_H /2$.
In order to account for the scale dependence at the Higgs vertex, the  
Sudakov form factor, $\sqrt{T_g(k_3,\mu_b)}$, will be included into 
the Green's function:
\beq
\label{green0}
G_0(\vk_2,\vk_3;x_1,x_2;\mu_b) =
\sqrt{T_g(k_3,\mu_b)}\; {k_2 ^4}\, \delta^{(2)}(\vk_2 - \vk_3)
\eeq
and
\beq
\label{greenbfkl}
G_{\mathrm{BFKL}}(\vk_2,\vk_3;x_1,x_2;\mu_b) =
 \, {k_2 k_3 \over 2\pi}\;
{\sqrt{T_g(k_3,\mu_b)} \over \sqrt{\pi D\log(x_1/x_2)}}\;
\exp\left[-{\log^2(k_2^2/k_3^2) \over 4D\log(x_1/x_2)}\right]
\, \left( x_1\over x_2\right) ^{\omega_0},
\eeq
where only the leading conformal spin was taken into account,
$\omega_0 = 4\log(2) \bar\alpha_s$ is the LL~BFKL intercept,
$D=14\zeta(3)\bar\alpha_s$ is the diffusion coefficient, and
$\bar\alpha_s = 3\alpha_s / \pi$.

\item The Higgs boson production vertex which does not differ from the
TPF case.

\item The off-diagonal unintegrated gluon distribution,
$f^{(b)}_g(x_3,\vk_3,\vq-\vk_3;\mu_b)$, of the gluon distribution 
that enters the Higgs boson production vertex from the lower side, 
with the total transverse momentum $\vq$. The gluon distribution 
is taken to be the same as in the case of the Two Pomeron Fusion, 
i.e.\ it is an off-diagonal gluon distribution given by
formulae \eq{kmrglue} and \eq{fqdep} and with $\mu_b = M_H /2$.

\item The off-diagonal unintegrated gluon distribution,
$f^{(c)}_g(x_4,\vk_4,-\vq-\vk_4;\mu_c)$,
for the screening Pomeron that propagates across the 
large rapidity distance from the lower proton to the triple Pomeron 
vertex with momentum transfer $-\vq$.
The scale $\mu_c$ is set by the gluon virtuality, so one has
$f^{(c)}_g(x_4,\vk_4,-\vq-\vk_4) = f^{(c)}_g(x_4,\vk_4,-\vq-\vk_4;k_4)$.
For a detailed discussion of this gluon distribution see Section~4.

\end{enumerate}

After these definitions the expression for the dominant imaginary part of 
the diagram of Fig.~\ref{Ampl2} can be written in the following form:
\[
\im M^{(1)}_{\mathrm{corr}}(y)
\, = \,
-\frac{9}{8}\; 16\pi^2 \; (2\pi^3 A)  \;
\int_{x_a} ^{x_b} \,
{dx_4 \over x_4}\,
\int\frac{d^2\,q}{(2\pi)^2}\,
\int\frac{d^2 k_2}{2\pi k_{2}^{4}}\,
\int\frac{d^2 k_3}{2\pi k_{3}^{4}}\,
\int\frac{d^2 k_4}{2\pi k_{4}^{4}}\,
\]
\[
\times \;
[V_{3P}\otimes f ^{(a)} _g(x_1)] (\vk_2,\,\vk_4)
G_{\mathrm{A}}(\vk_2,\vk_3;x_1,x_2; M_H/2)\,
\]
\beq
\times\;
f^{(b)}_{g}(x_3,\vk_3,\vq-\vk_3; M_H/2)\,
f^{(c)}_{g}(x_4,\vk_4,-\vq-\vk_4),
\label{corr}
\eeq
where $\frac{9}{8} = \frac{N_c^2}{N_c^2-1}$ 
tends to unity in the large $N_c$ limit, 
and the coefficient $\,2\pi^3\,A\,$ has been extracted in order 
to enable an easier comparison with the Two Pomeron Fusion contribution 
given by \eq{born}.
$x_a$ and $x_b$ denote the lower and the upper limits of the 
integration over $x_4$, respectively. 
The fractions of the longitudinal momenta 
of the upper proton $x_1$ and $x_2$ and of the lower proton $x_3$, $x_4$ read:
\beq
x_3 =  {M_H \over \sqrt{s}} \, \exp(-y), \qquad
x_2 = {M_H \over \sqrt{s}} \, \exp(y), \qquad
x_1 \simeq {k_0^2 \over  x_4 s}
\label{xdef1}
\eeq
\beq
x_a< x_4 <x_b,\qquad
x_a \simeq {k_0^2 \over  s},\qquad
x_b \simeq {k_0^2 \over  x_2 s}.
\label{xdef2}
\eeq
The energy scale $k_0^2$ that enters the definitions of gluon~$x_i$ 
corresponds to typical values of the virtualities, and we set
$k_0^2 = 1$~GeV$^2$. The dependence on the momentum transfer~$\vq$,
circulating in the Pomeron-proton triangle, appears only inside the off-shell 
unintegrated gluon distributions, $f_g ^{(b)}$ and $f_g ^{(c)}$, 
which contain the rapidly varying proton form-factor. 
The $\vq$~dependence of the other vertices and of the Green's 
function, $G_A$, is weak and will be neglected.
The amplitude (\ref{corr}) includes a factor of~2 
coming from the possible emissions of the Higgs boson from both of
the Pomerons below the triple Pomeron vertex. In addition, we have to add 
another contribution in which we interchange the role of {\em projectile} and
{\em target}: since the structure of both protons is identical, we simply 
replace $y \to -y$. So finally the correction amplitude may be written as:
\beq
M_{\mathrm{corr}}(y) = M^{(1)} _{\mathrm{corr}}(y) +  
M^{(1)} _{\mathrm{corr}}(-y).
\label{corrtot}
\eeq
In the case of the central production, $y=0$,
and $\; M_{\mathrm{corr}}(y) = 2\,M^{(1)}_{\mathrm{corr}}(y)$.

\subsection{Infrared stability}

It is instructive to study the infrared behaviour of the integral
in \eq{corr} using a simplified framework and the saddle point method.
First of all, we note that the integrals in \eq{corr} are infrared finite.
This follows from properties of the triple Pomeron vertex (which vanishes 
whenever one of the external momenta goes to zero) and of the 
unintegrated gluon densities (which, as a result of gauge invariance, also 
vanish as one of the two gluon momenta goes to zero). 
Next, after averaging over the angles in $\vk_2$ 
and $\vk_4$, the triple Pomeron vertex (\ref{Triple}) imposes the 
conditions $k_2 \geq k_1$ and $k_4 \geq k_1$.
Finally, disregarding irrelevant constants and details one finds that, for the 
real contribution of the triple Pomeron vertex, the correction term 
(\ref{corr}) behaves as
\beq
Q_R \sim
\int_{k_0^2}\frac{dk^2_1}{k_{1}^{2}}\,
\int_{k_1^2}\frac{dk^2_2}{k_{2}^{4}}\,
\int_{k_1^2}\frac{dk^2_4}{k_{4}^{4}}\,
f^{(a)}_{g}(x_2,k_1^2)\,
f^{(b)}_{g}(x_3,k_2^2; M_H/2)\,
f^{(c)}_{g}(x_4,k_4^2)\,
\sqrt{T_s(k_2\,,M_H/2)},
\label{crude}
\eeq
where we used, for the Green's function $G_A$, the approximation of 
the two-gluon exchange, see \eq{green0}. Note that the Sudakov form-factor,
$\sqrt{T_s(k_2\,,M_H/2)}$, is present in the Green's function 
(shown explicitly in (\ref{crude})) and, implicitly, in the gluon 
distributions $f^{(b)}_{g}(x_3,k_2^2; M_H/2)$.

In this analysis we have introduced a cut-off, $k_0$, of the low
momenta in the integration over $k_1$. The scale $k_0$ is meant to lie in the 
perturbative domain above the low hadronic scale of the proton form-factor.
The structure of the nested integrations in (\ref{crude}) implies that this 
scale $k_0$ is a lower limit for all momenta. 
By studying the variation of the integrals  
(\ref{crude}) as a function of $k_0$, we obtain a measure of the sensitivity 
of the correction to poorly known details of the proton structure at low 
momentum scales. The strong coupling constant, $\alpha_s$, is kept fixed, 
as the running coupling introduces only logarithmic corrections, 
and, for the moment, we are interested
in tracing only the leading power of $k_0$ that arises from the integral 
(\ref{crude}). In our estimate we use the Sudakov form-factor in the 
double logarithmic approximation:
\beq
T_s(k,\mu) = \exp \biggl( -{3\alpha_s \over 4\pi} \log^2 (\mu^2 / k^2) \biggr).
\eeq

Let us start by performing the analysis of $Q_R$ with the conservative 
assumption of a vanishing anomalous dimension of the gluon density (i.e. a 
constant behaviour of the gluon density as a function of momentum).
Then the integral over $k_2$ behaves in the same way as the 
integral defining the Two Pomeron Fusion amplitude. 
It is dominated by a contribution from a saddle point
at a perturbative value, $k_2 = (M_H/2)\exp(-2\pi\, / \, 3\alpha_s )$,
with only marginal sensitivity to the infrared domain.

The nested integrations over $k_1$ and $k_4$ in \eq{crude} are dominated by
momenta close to the  cut-off scale $k_0$ and lead to a result
$\sim 1/k_0^2$. This means that our correction is not small, but the 
magnitude is sensitive to details of the gluon distributions 
$f_g^{(a)}$ and $f_g^{(c)}$ at low momenta, and  hence, unreliable. At first 
sight, this seems to imply that the whole Ansatz of choosing, for the 
absorptive correction, a perturbative gluon ladder, cannot be justified.

A closer look, however, shows that this may not be the final answer.   
Namely, many arguments have been given that the gluon density at very small 
$x$ should saturate, i.e. it scales and depends upon the ratio 
$k^2/Q^2_s(x)$. Here $Q_s(x) \sim x^{-\lambda}$ is the saturation scale 
which grows in $1/x$ with $\lambda \simeq 0.3$. This leads to a substantial 
change of the $k^2$-dependence of the gluon density for momenta being 
smaller than the saturation scale: the saturation scale becomes an effective
lower cut-off of the gluon momenta in the small-$k^2$ region.
The phenomenological analysis of 
HERA data~\cite{GBW1,GBW2} indicates that $Q^2_s(x_4)$ takes 
values of a few GeV$^2$ for the range of $x_4$ relevant in our problem 
where $\log(1/x_4)$ can be as large as 14~to~19 units.  
 
Applying these arguments to our integral \eq{crude} we find that, in the    
presence of the saturation scale, the momentum dependence of gluon 
distribution $f_g^{(c)}$ is modified, and the nested integrals 
over $k_1$ and $k_4$ give
\beq
\int_{Q_s^2(x_4)}\frac{dk^2_4}{k_{4}^{4}}
\int_{k_0^2} ^{k_4^2}
\frac{dk^2_1}{k_{1}^{2}
}\, \sim
{\log(Q_s^2(x_4) / k^2_0) \over Q_s^2 (x_4)}.
\eeq
With $Q_s(x_4)$ being well above the infrared scale $k_0^2$, this
represents a well-behaved result with a low sensitivity to the 
infrared physics. Keeping in mind that, in \eq{crude}, the integration 
over $k_2$ is infrared-safe, one finds that $Q_R$ is infrared stable.

An analogous reasoning can be applied also
to the contribution of the virtual correction part of
the triple Pomeron vertex to the integrals in \eq{corr}.
The behavior is approximated by:
\beq
Q_V \sim
\int_{k_0^2}\frac{dk^2_1}{k_{1}^{4}}\,
\int_{k_1^2}\frac{dk^2_2}{k_{2}^{4}}\,
\log(k_2^2/k_1^2)
f^{(a)}_{g}(x_2,k_1^2)\,
f^{(b)}_{g}(x_3,k_1^2; M_H/2)\,
f^{(c)}_{g}(x_4,k_2^2)\,
\sqrt{T_s(k_1\,,M_H/2)},
\label{crudev}
\eeq
and a similar term, $Q_V '$, is obtained by interchanging
$f^{(b)}_{g}(x_3,k_1^2;M_H/2) \to f^{(b)}_{g}(x_3,k_2^2; M_H/2)$ and
$f^{(c)}_{g}(x_4,k_2^2) \to f^{(c)}_{g}(x_4,k_1^2)$.
One finds that the integral over $k_2$ ($k_1$) in $Q_V$ ($Q_V '$)
is stabilized in the infrared by the saturation scale, and the integral
over $\vk_1$ ($\vk_2$) is stabilized by the  Sudakov form-factor.

Collecting the estimates for $Q_R$, $Q_V$ and $Q'_V$
one concludes that the perturbative determination of the
amplitude of screening correction is possible provided that the gluon 
distribution $f^{(c)}_{g}(x_4,k_4^2)$ possesses the saturation property, 
i.e.\ the simple gluon ladder that we have used initially has to be corrected.
In this case the momentum scale at the upper end of the screening 
correction, $f_g^{(c)}$, is not small, and our perturbative Ansatz 
is justified. Note that, in this context, saturation plays a role quite 
analogous to the Sudakov form factor at the production vertex:
without the Sudakov form factor the momentum integral near the production 
vertex would not have been infrared-safe. Similarly, saturation renders the 
nested $k_1$ and $k_4$ integrations infrared-safe.  
    
\begin{figure}[t]
\begin{minipage}{12.0 cm}
\begin{center}
\epsfxsize=10cm
\leavevmode
\hbox{ \epsffile{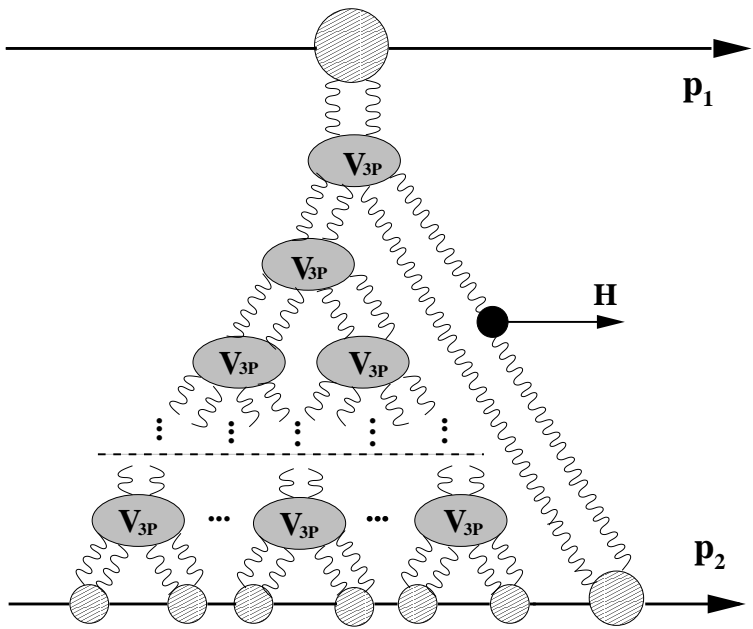}}
\end{center}
\end{minipage}
\caption{\it The fan diagram approximation to the hard rescattering
correction  to the exclusive Higgs boson production.}
\label{Ampl4}
\end{figure}

\section{Nonlinear evolution of the gluon distribution}

In this section we use the simplest available tool, the nonlinear Balitsky-
Kovchegov (BK) equation~\cite{Bal,Kov1,Kov2}, in order to model the 
unitarity corrections for the screening Pomeron ladder in Fig.~\ref{Ampl2}.   
This equation resums fan diagrams of BFKL Pomerons (Fig.~\ref{Ampl4}) in the 
LL$1/x$~approximation, and it has been derived, in the large $N_c$ limit, for
the case of a small probe scattering off a large nucleus. The BK equation
generates, in a natural way, the saturation scale growing exponentially
with rapidity~\cite{MB,GBMS}. 
The equation has originally been derived for the dipole scattering amplitude, 
but later on it was re-formulated as an explicit
equation for the evolution of the unintegrated gluon 
distribution~\cite{KK,KS}, and in this paper we employ the latter form.

The hard rescattering correction amplitude in \eq{corr} was given in
the momentum representation, in order to match the variables of the 
triple Pomeron vertex. In this section, however, we find it more convenient 
to do a partial Fourier transform from momentum transfer carried
by the Pomerons to the impact parameter variable.
The impact parameter space seems to be advantageous, both in performing
the non-linear gluon density evolution and for evaluating of the soft 
gap survival factor.
The correct treatment of the impact parameter dependence of the scattering
amplitude is difficult in the BK framework. 
In the original formulation
of the BK equation, the gluons are assumed to be massless, in accordance
with perturbative QCD; this contradicts, however, essential features of 
non-perturbative QCD dynamics leading to the confinement of colour.
In particular, in the perturbative framework one finds contributions 
from large dipoles which
generate a power like tail of the impact parameter profiles~\cite{KOVN1,GBS}.
Such a tail leads to a violation of the Froissart bound~\cite{KOVN2}.
Therefore, the limit of large impact parameters of the BK must be
modified, using some phenomenological model. 
In practice, all the applications of the BK
equation to the proton structure are based on the approximation of
treating the evolution at fixed impact parameter~\cite{GBMS,KK,GLLM,IIM}, 
i.e. the evolution proceeds independently at each value of impact 
parameter~$\vb$. 

Thus, our starting point is the form of BK equation proposed in
Ref.~\cite{KK}:
\[
\frac{\partial f_g (x,k^2)}{\partial \log 1/x}=  
\frac{\alpha_s N_c}{\pi} \,k^2\int_{k_0^2}{dk^{\prime 2}\over
k^{\prime 2}}\left \{{f_g \left({x},k^{\prime 2}\right)-
 f_g \left({x},k^2\right)\over |k^{\prime 2}-k^2|} +
{f_g\left({x},k^2\right)\over [4k^{\prime 4}+k^4]^{{1\over 2}}}\right\}
\]
\begin{equation}
-\alpha_s^2\, \left(1-k^2{d\over dk^2}\right)^2{k^2\over R^2}
\left[\int_{k^2}^{\infty}
{dk^{\prime 2}\over k^{\prime 4}}\, \log\left(
{k^{\prime 2}\over k^2}\right)
\,
f_g(x,k^{\prime 2})\right]^2 \; ,
\label{eq:bk}
\end{equation}
where the assumption has been made that the distribution of gluons 
is uniform inside a transverse disc with the radius $R$. The 
unintegrated gluon density in the impact parameter plane therefore 
takes the form:
\beq
\tilde f_g(x,k^2,b) = f_g(x,k^2) \; {\theta(R-b) \over \pi R^2}.
\eeq
In (\ref{eq:bk}), the factor $1/R^2$ in front of the non-linear term is a 
result of the impact parameter integration:
\beq
\int d^2 b \, \tilde f_g(x,k_1^2,b) \tilde f_g(x,k_2^2,b) = {1\over \pi R^2}\;
f_g(x,k_1^2) \, f_g(x,k_2^2).
\eeq
In our case, however, it is necessary to control the impact
parameter dependence in more detail. 
Therefore we assume, at $x_0 = 0.01$, an initial Gaussian impact parameter 
profile: 
\beq
\tilde f_g (x_0,k^2,b) = f_g(x_0,k^2)\, {\exp(-b^2 / R^2) \over \pi R^2},
\eeq
and we set the radius $R \simeq 2.83$~GeV$^{-1}$ ($R^2 = 8$~GeV$^{-2}$), 
in accordance with  the proton radius seen in the $t$-dependence 
of the exclusive $J/\psi$ production at HERA. 
The evolution equation for the density of the unintegrated gluon 
reads:
$$  
\frac{\partial \tilde f_g(x,k^2,b)}{\partial \log 1/x}=  
\frac{\alpha_s N_c}{\pi} \,k^2\int_{k_0^2}{dk^{\prime 2}\over  
k^{\prime 2}}
\left \{
{\tilde f_g\left({x},k^{\prime 2},b\right)-
\tilde f_g\left({x},k^2,b\right)\over |k^{\prime 2}-k^2|} +
{\tilde f_g\left({x},k^2,b\right)\over 
[4k^{\prime 4}+k^4]^{{1\over 2}}}\right\}
$$
\begin{equation}  
-\pi \alpha_s^2\, \left(1-k^2{d\over dk^2}\right)^2{k^2}  
\left[\int_{k^2}^{\infty}  
{dk^{\prime 2}\over k^{\prime 4}}\, 
\log\left(  
{k^{\prime 2}\over k^2}\right) \,
\tilde f_g (x,k^{\prime 2},b)\right]^2 \; .
\label{eq:bkb}  
\end{equation} 
This evolution is local in~$\bb$, but due to the 
non-linear term, for $x<x_0$, the impact parameter dependence cannot be 
factorized out from the resulting unintegrated gluon density, 
$\tilde f_g(x,k^2,b)$.

It is clear, however, that the BK equation (\ref{eq:bk}) needs many 
improvements to be applicable in phenomenology, in particular it is 
essential to take into account, in some approximate manner, the NLL$1/x$ 
corrections to the linear BFKL kernel~\cite{nlbfkl1,nlbfkl2}.  
Following  the path proposed in series of papers~\cite{KMS,KK,KS},
we include, in the linear evolution kernel, the following improvements:
\begin{itemize}
\item the consistency constraint~\cite{cclund,ccdur} 
that models an essential part of the NNL$1/x$ correction to the BFKL 
kernel and realizes, approximately, also a resummation of the corrections to 
all orders; 
\item the non-singular part in the $x \to 0$ limit of the DGLAP splitting
function is added;
\item the DGLAP quark density evolution is also taken into account,  
and the evolution equation is coupled to the gluon density evolution using
the DGLAP splitting functions;
\item the strong coupling constant is running with the scale given by
gluon virtualities in the BFKL ladder.

\end{itemize}
These modifications are consistent with the exact NLL$1/x$ corrections
to the BFKL kernel, and they impose the constraints on the evolution which 
follow from the DGLAP formalism in the collinear limit \cite{col1,col2,col3}.
The implementation of these corrections has been described in detail in the 
recent papers \cite{KK,KS}, and the final form of the equation
can be found there as well. It is rather straightforward to add, 
on top of the modifications described above, the impact parameter 
dependence, in a manner very similar to \eq{eq:bkb}.
The normalisation of the input parameterisation for the gluon
and quark distributions at high~$x$ has been tuned to provide a good fit
of the $F_2$ data \cite{KUTM}, and the obtained distributions have been 
shown to give a good description of inclusive heavy quark production
at HERA and at the Tevatron \cite{JKMP}.  
A detailed description of the properties of the obtained partonic
densities will be given in a separate paper \cite{KUTM}. 

In our numerical calculations of the exclusive Higgs boson production 
(\ref{corr}) we will take $\tilde f^{(c)}(x,k^2,b)$ as being the solution 
to the collinearly improved, impact parameter dependent BK equation. 
Let us remind that the other two  
gluon distributions of \eq{corr}, $\tilde f^{(a)}(x,k^2,b)$ and  
$\tilde f^{(b)}(x,k^2,b)$,  are probed at rather large values 
of $x \sim 0.01$, and non-linear effects can safely be neglected. Instead, 
we apply the prescriptions (\ref{unglue}) and (\ref{kmrglue}), 
using with the CTEQ5 parameterisation for the collinear gluon distribution 
functions.

\section{Soft gap survival probability}

Thus far, we have determined the hard component of the amplitude for exclusive
Higgs boson production. The condition to leave the protons intact in the
production process, however, imposes constraints on the soft rescattering
of the protons. The soft gap survival probability factor, $\hat S^2$,
quantifies the probability that the protons emerge from the reaction intact
despite possible soft rescatterings. The gap survival probability can
conveniently be estimated in the impact parameter representation by employing,
for instance, the two-channel eikonal model~\cite{KMR11,KMR12}. 
Therefore, we now translate the exclusive Higgs boson production cross 
section into the impact parameter representation.

Let us start with a general discussion. We consider
a scattering amplitude of an exclusive production in the impact parameter
representation
\beq
\tilde M(\vb_1,\vb_2) =
\int {d^2 q_1 \over (2\pi)^2} \int {d^2 q_2 \over (2\pi)^2}
\, M(\vq_1,\vq_2) \, e^{i\vqq_1 \vbb_1} \, e^{i\vqq_2 \vbb_2},
\eeq
where $\, M(\vq_1,\vq_2)\,$ is the amplitude given as a function of the
transverse momenta. Then, the integrated cross-section
\beq
\sigma = {1\over 16\pi} \,
\int {d^2 q_1 \over (2\pi)^2} \int {d^2 q_2 \over (2\pi)^2}
|M(\vq_1,\vq_2)|^2,
\eeq
can be expressed in the following way:
\beq
\sigma = {1\over 16\pi} \,\int {d^2 b_1} \int {d^2 b_2} \;
|\tilde M(\vb_1,\vb_2)|^2,
\eeq
or, alternatively,
\beq
\sigma = {1\over 16\pi} \,\int {d^2 b_1} \int {d^2 b} \;
|\tilde M(\vb_1,\vb-\vb_1)|^2,
\eeq
where $\vb_1$ and $\vb_2$ are transverse positions of the production vertex 
measured from centres of the protons, and $\vb$ is the impact parameter 
of the collision.

\begin{figure}[t]
\begin{center}
\epsfxsize=14.5 cm
\leavevmode
\hbox{\epsffile{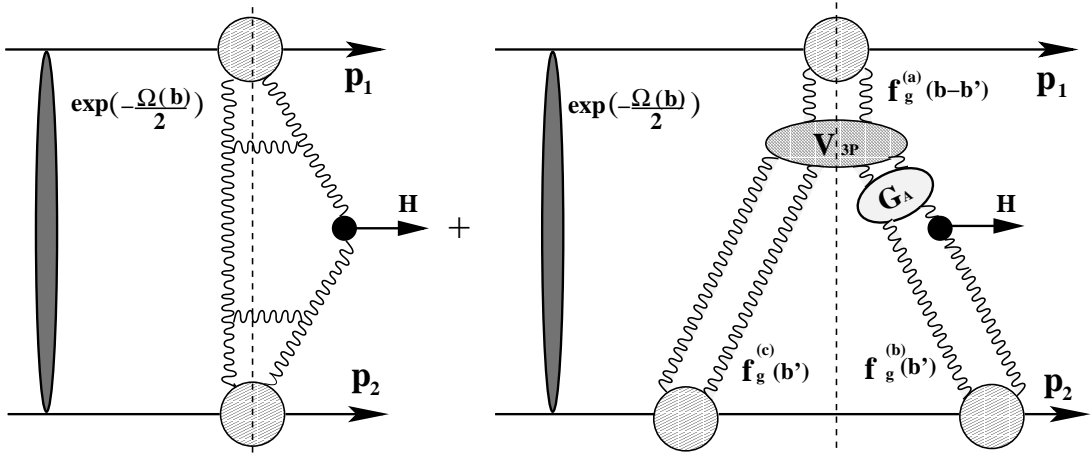}}
\end{center}
\caption{\it 
The exclusive $pp \to p\;H\;p$ amplitude with a hard and soft 
rescattering correction.}
\label{figsoft}
\end{figure}

In the Two Pomeron Fusion case, the off-diagonal unintegrated gluon
distributions are probed at moderate values of $x \sim 0.01$ and at
virtualities $k^2 > 1$~GeV$^2$, so one can assume the gluon density 
to take a factorized form:
\beq
\tilde f_g ^{\mathrm{off}}(x,x',k^2,b;\mu) = 
f_g ^{\mathrm{off}}(x,x',k^2;\mu)\,S(b)
\eeq
with
\beq
\label{imp}
S(b)\,=\,\frac{1}{\pi\,R^{2}}\,e^{-\frac{b^2}{\,R^{2}}}\,,
\eeq
where $\,R\,\simeq \,2.83\,$~GeV$^{-1}\,$ is the proton radius, measured 
in the exclusive production of $J/\psi$ production at HERA
(cf. our discussion of the impact parameter dependent BK~equation),
and $S(b)$ is normalized
\beq
\int\,d^2\,b\,S(b)\,=\,1\,\,.
\eeq
Therefore, the cross section for the production of the Higgs boson
in the Two Pomeron Fusion approximation takes the form:
\beq
{d\sigma^{(0)}_{pp\to pHp}(y) \over dy} = {1\over 16\pi}\,
\int d^2 b \int d^2 b_1 |S(\vb_1)S(\vb-\vb_1)M_0(y)|^2,
\label{sigmab}
\eeq
with $M_0(y)$ given by \eq{born}.
Including the hard rescattering correction, we arrive at the following
expression:
\beq
{d\sigma^{(0+1)}(y)_{pp\to pHp}  \over dy} = {1\over 16\pi}\,
\int d^2 b \int d^2 b_1 |S(\vb_1)S(\vb-\vb_1)M_0(y)+
\tilde M_{\mathrm{corr}}(y,\vb,\vb_1)|^2,
\eeq
where the dominant parts of both $M_0(y)$ and 
$\tilde M_{\mathrm{corr}}(y,\vb,\vb_1)$ are imaginary, and
\[
\im \tilde M_{\mathrm{corr}}(y,\vb,\vb_1)
\, = \,
-\left\{
\frac{9}{8}\; 16\pi^2 \;(2\pi^3 A)  \;
\int_{x_a} ^{x_b} \,
{dx_4 \over x_4}\,
\int\frac{d^2 k_2}{2\pi k_{2}^{4}}\,
\int\frac{d^2 k_3}{2\pi k_{3}^{4}}\,
\int\frac{d^2 k_4}{2\pi k_{4}^{4}}\,
\right.
\]
\[
\times\;
[V_{3P}\otimes \tilde f ^{(a)} _g(x_1,\vb-\vb_1)] (\vk_2,\,\vk_4)\,
G_{\mathrm{A}}(\vk_2,\vk_3;x_1,x_2; M_H/2)\,
\]
\beq
\left.
\times\;
\tilde f^{(b)}_{g}(x_3,k_3^2, \vb_1; M_H/2)\,
\tilde f^{(c)}_{g}(x_4,k_4^2,\vb_1) 
\right\} \; - \; \{ y \to -y\}.
\label{bcorr}
\eeq
The impact parameter dependence of
$\tilde f_g ^{(a)}$ and $\tilde f_g ^{(b)}$ is given by $S(b)$ and may be
factorized out, in analogy with the case of the Two Pomeron Fusion amplitude.
The impact parameter dependence of $\tilde f_g ^{(c)}$ is determined by the
BK~evolution, and it is different from $S(b)$.
Note that, in \eq{corr}, the convolution in the momentum transfer~$\vq$  
leads to a product of~$\tilde f_g ^{(b)}$ and
$\tilde f_g ^{(c)}$, evaluated at the same point, $\vb_1$, in the transverse 
position space.

Having defined $\tilde M_{\mathrm{corr}}(y,\vb,\vb_1)$, it is
straightforward to write down the cross section for the exclusive
production, taking into account the soft rescattering
\beq
{d\sigma^{(0+1),\Omega} _{pp\to pHp} (y) \over dy} = {1\over 16\pi}\,
\int d^2 b \int d^2 b_1\,
|S(\vb_1)S(\vb-\vb_1)M_0(y)+M_{\mathrm{corr}}(y,\vb,\vb_1)|^2\,
\exp(-\Omega(s,\,\vb)),
\label{final}
\eeq
where $\,\Omega(s,\,\vb)\,$ is the opacity of the $\,pp\,$ scattering
at squared energy $s$ and impact parameter $\vb$.
In this paper we adopt the opacity factor of 
the two channel eikonal model proposed in \cite{KMR11,KMR12}.
The gap survival probability reads:
\beq\
\label{sp}
\hat{S}^2 =
\frac{\left(\,
\frac{d\sigma_{pp\to pHp}^{(0+1),\Omega} (y)}{dy}
\, \right)_{y=0}
}
{
\left(\,
\frac{d\sigma_{pp\to pHp}^{(0+1)}}{dy}
\, \right)_{y=0}
} \; .
\eeq
The impact parameter profile of the rescattering correction amplitude
(\ref{bcorr}) is different from the Two Pomeron Fusion case; thus the
gap survival factor will take different values in those two cases.

\section{Results}

We have performed all numerical estimates of the amplitude and of the cross 
section for the central exclusive Higgs boson production, choosing y=0, 
assuming the LHC energy, $\sqrt{s} = 14$~TeV, and taking, for the  Higgs 
boson mass, $M_H =120$~GeV. 
As we have stated before, in this paper our main goal is a general 
understanding of the hard screening mechanism and an estimate of its 
magnitude, rather than a precise determination of the cross section.

As a first step, we have considered the rescattering due to a single BFKL 
ladder. As expected, due to the long rapidity evolution of the BFKL 
amplitude this correction is large. As we have discussed  
before, since this single-ladder correction is highly sensitive to the 
infrared region, the precise numerical value of its magnitude is not 
reliable. From this part of our analysis, we therefore only conclude that 
this type of hard rescattering may potentially be large and, in any realistic 
estimate, unitarity corrections of the screening BFKL ladder have 
to be taken into account.

In the next step of our numerical analysis, therefore, we have made use 
of the BK-corrected unintegrated gluon density, $f_g^{(c)}$. 
First, using \eq{bcorr}, we have evaluated the hard 
rescattering amplitude as a function of the impact parameter, at fixed 
$x_4$. The contribution of momenta lower than $k_0=1$~GeV was assumed 
to vanish. The main reason for that is that the gluon densities are
poorly known for gluon momenta lower than $k_0$, so is the behaviour 
of gluon propagators, and in that region one has to rely on uncertain 
extrapolations. On the other hand, with the integrand of \eq{corr} 
being positive, the application of the cut-off leads to 
a conservative estimate of the correction term. Still, the main
contribution to the correction amplitude comes from the range of momenta 
around the saturation scale $Q_s(x_4)$.

Recall that, 
for the four-point gluon Green's function which connects the Higgs 
vertex with the triple Pomeron vertex, we consider two different scenarios 
-- the simple two gluon exchange (\ref{green0})
and the BFKL Green's function in the diffusion approximation 
(\ref{greenbfkl}). 
The strong coupling constant was fixed at $\alpha_s = 0.15$, both 
in the Green's function and in the triple Pomeron vertex. 
Such a choice is suitable for the LL~BFKL Green's function, 
as the small~$\alpha_s$ reduces both the Pomeron intercept and 
the diffusion coefficient, in line with the NLO corrections 
to the BFKL evolution. This value of $\alpha_s$ seems, however, to 
be too small for the triple Pomeron vertex: thus the estimate of
the correction is conservative. Therefore, in the vertex, we shall consider 
also values of $\alpha_s$ other than our default choice 
of $\alpha_s = 0.15$, while in the BFKL Green's function we always keep 
$\alpha_s=0.15$.

\begin{figure}[t]
\begin{center}
\epsfxsize=11.5 cm
\leavevmode
\hbox{ \epsffile{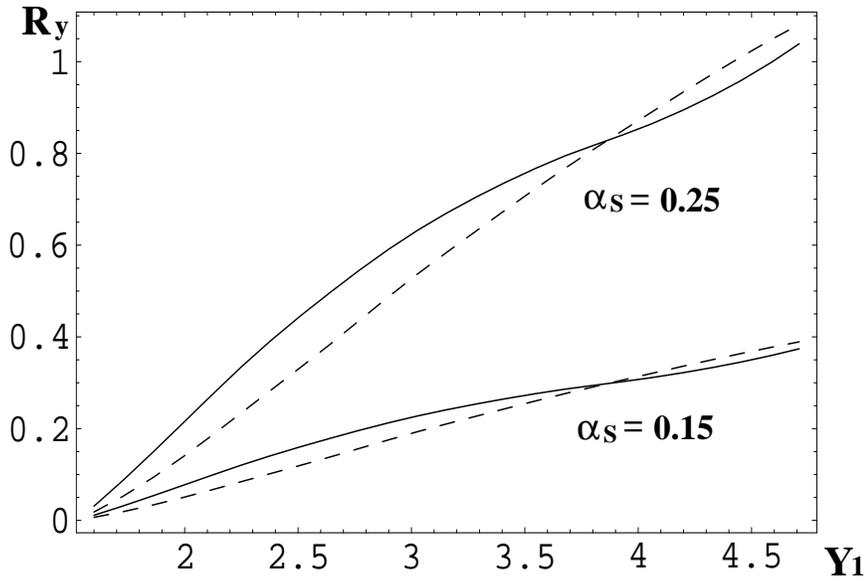}}
\end{center}
\caption{\it The ratio $R_y(y_1)$ (see \eq{ry}) 
of the hard rescattering correction to the Two Pomeron Fusion
term at $b=0$, as a function of the rapidity distance, $y_1$, of
the Triple Pomeron Vertex from the {\em projectile} protons, for
$\,\alpha_s\,=\,0.15\,$ and $\,\alpha_s\,=\,0.25\,$ at 
$\sqrt{s}=14$\,\,TeV.
The black line belongs to the Green's function, $G_A = G_{\mathrm{BFKL}}$,
the dashed line to the approximation of two elementary gluons, $G_A=G_0$.}

\label{Ampl5}
\end{figure}

In Fig.~\ref{Ampl5} we present the total hard rescattering 
correction to the amplitude in the form of the ratio 
\beq
R_y (y_1) \; = \; \left[{|d\tilde M_{\mathrm{corr}}(y,\vb,\vb_1) / dy_1| 
\over |\tilde M_0(y,\vb,\vb_1)| }
\right]_{b=b_1=0,\, y=0},
\label{ry}
\eeq
with the impact parameter dependent Two Pomeron Fusion amplitude
being given by (cf.\ \eq{born} and \eq{sigmab})
\beq
\tilde M_0(y,\vb,\vb_1) = M_0(y)\, S(\vb-\vb_1) S(\vb_1).
\eeq 
Here $M_0(y)$ has also been evaluated with the
infrared cut-off $k_0=1$~GeV, and the off-diagonal unintegrated gluon 
distributions have been obtained from the CTEQ5 collinear gluon distribution. 
Note that we plotted the ratio as a function of the rapidity related 
variable, $y_1 = \log (1/x_1)$,  instead of the less convenient 
variable~$x_4$ (the relation between these variables is given by~\eq{xdef1}). 
The variable $y_1$ has the meaning of a rapidity distance of the triple 
Pomeron vertex from the {\em projectile} proton, and it is related to the 
invariant mass of the rescattering intermediate state by  
$M_X^2 \sim k_0^2\, \exp(y_1)$.

In Fig.~\ref{Ampl5} we compare the results, obtained with the two choices
of Green's function  and with the two values $\alpha_s=0.15$ and 
$\alpha_s=0.25$. 
The relative sign of the correction term is negative. 
From this figure we draw the following conclusions:

\begin{itemize}

\item 
The hard rescattering correction is the highest for the highest 
mass of the diffractive intermediate state. 
This is due to the strong suppression of the gluon density in the 
proton at large values of the gluon~$x$. We interpret this as an 
indication that the contribution of the hard rescattering correction
is clearly separated from the soft rescattering contributions of the
two-channel eikonal model~\cite{KMR11,KMR12} which incorporates only
the low mass diffractive intermediate states.

\item 
There is no significant difference between the BFKL Green's function 
and the exchange of two elementary gluons. The main reason is that 
the BFKL evolution length is short for larger values of $y_1$, which 
give the dominant part of the corrections. 
The relative BFKL effects are more pronounced for small $y_1$,
but this region provides only a small contribution.

\item 
The relative magnitude of the correction is quite significant for 
$\alpha_s=0.15$, and for $\alpha_s=0.25$ the correction even exceeds 
the Two Pomeron Fusion amplitude. The interpretation of this striking result 
will be discussed in the next section.

\end{itemize}

\begin{figure}[t]

\begin{center}
\epsfxsize=11.5 cm
\leavevmode
\hbox{ \epsffile{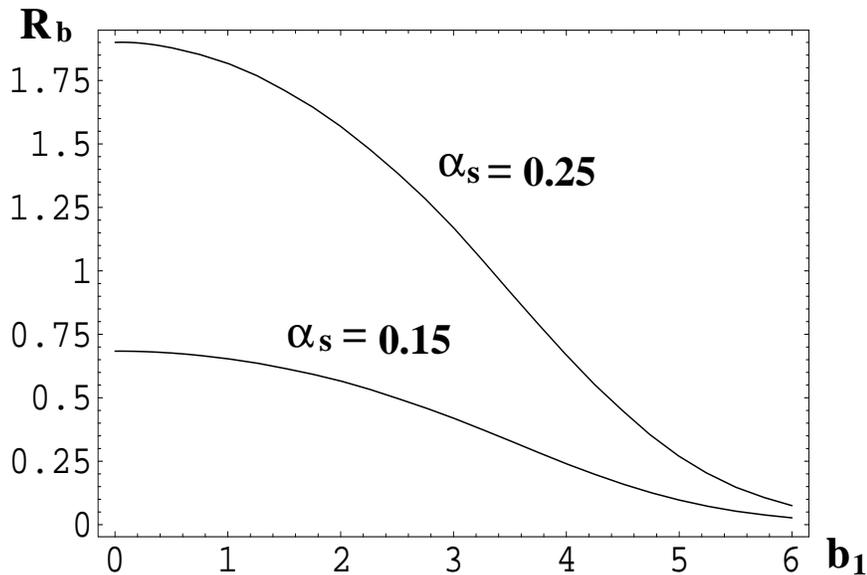}}
\end{center}
\caption{\it The ratio $R_b(b_1)$ (see \eq{rb}) 
of the hard rescattering correction to the Two Pomeron Fusion
term at $b=0$, as a function of the transverse distance, $b_1$ of
the Triple Pomeron Vertex from the centre of the {\em target} proton, for 
the two values $\,\alpha_s\,=\,0.15\,$ and $\,\alpha_s\,=\,0.25\,$ at 
$\sqrt{s}=14$\,\,TeV. We only show the results for the BFKL Green's function.}
\label{Ampl6}
\end{figure}

In Fig.~\ref{Ampl6}, results are shown for the impact parameter 
dependent correction amplitude, integrated over $y_1$, as a function
of the transverse distance, $\vb_1$, of the Higgs boson production vertex 
from the center of the {\em target} proton:  
\beq
R_b (b_1) \; = \;
\left[|\tilde M_{\mathrm{corr}}(y,b,b_1)| \over 
|\tilde M_0(y,b,b_1)| 
\right] _{y=0, \, b=0} \, .
\label{rb}
\eeq
Here we have set $b=0$, but, in fact, in the ratio the impact parameter 
dependence coming from the form factor of the {\em projectile} proton 
drops out,
and any value of~$b$ can be taken. In Fig.~\ref{Ampl6} only the results 
obtained with the BFKL Green's function are displayed. There is not much 
difference between the BFKL Green's function and the two gluon approximation.
One sees in the figure that 
the ratio of the absorptive correction profile, 
$\tilde M_{\mathrm{corr}}(y,b=0,b_1)$,  
to the Two Pomeron Fusion profile, $\tilde M_0(y,b=0,b_1)$, decreases with 
increasing distance, $b_1$, from the proton center. Thus, the correction
term tends to suppress the Higgs production inside the {\em target} proton. 
It is also clearly visible 
that the correction is large for both choices of $\alpha_s$.


\begin{table}
\begin{center}
\begin{tabular}{|c|c|c|c|c|c|}
\hline
\, & \, & \,& \,& \, & \, \\
$d\sigma_{\mathrm{TPF}}/dy$ & $\,d\sigma/dy\,(\alpha_s\,=0.15)\,$ &
$\,d\sigma/dy\,(0.17)\,$ & $\,d\sigma/dy\,(0.2)\,$ &
$\,d\sigma/dy\,(0.22)\,$ & $\,d\sigma/dy\,(0.25)\,$ \\
\, & \,  & \,& \,& \, & \,\\
\hline
\, & \, & \,& \,& \, & \, \\
0.4 & 0.16 & 0.12 & 0.042 & 0.08 & 0.14 \\
\, &  \, & \,& \,& \, & \, \\
\hline
\end{tabular}
\caption{\it The differential cross section 
$\,\Le\frac{d\sigma_{pp\to pHp}}{d\,y}\Ra_{y=0}\,$ (in fb),
obtained from the Two Pomeron Fusion amplitude, and corrected by the 
hard rescattering and by the soft gap survival probability
factor, for various values of $\,\alpha_s\,$ at $\sqrt{s}=14\,\,TeV\,$.}
\label{Ampl7}
\end{center}
\end{table}

Performing, in \eq{final}, the integrations over the transverse positions, 
~$\vb_1$ and~$\vb$, leads to the differential 
cross section of the exclusive Higgs boson production, 
${d\sigma^{(0+1),\Omega} _{pp\to pHp} (y)\,  / \, dy}$ which takes 
into account both the soft and the hard rescattering corrections. 
The results of this calculation, for various values of~$\alpha_s$ in
the triple Pomeron coupling, are collected in Tab.~\ref{Ampl7}.
At $\alpha_s \simeq 0.2\,$ the absolute value of the hard rescattering 
correction exceeds the Two Pomeron Fusion term, and this explains 
why, at this value of $\alpha_s$, the corrected cross section takes its 
smallest value.  For all values of $\alpha_s$ that we have considered 
the cross section with the hard rescattering correction is smaller than 
the Two Pomeron Fusion cross section by factor larger than~2.5.
Let us repeat that, by varying $\alpha_s$, 
we want to illustrate the sensitivity of the result to the size of the 
triple Pomeron coupling. Further discussion will be given in the next 
section.

\begin{table}[t]
\begin{center}
\begin{tabular}{|c|c|c|c|c|c|}
\hline
\, & \, & \,& \,& \, & \, \\
$\hat{S}^{2}\,\,(\mathrm{TPF})\,\,$ & $\,\hat{S}^{2}\,(\alpha_s\,=0.15)\,$ &
$\,\hat{S}^{2}\,(0.17)\,$ & $\,\hat{S}^{2}\,(0.2)\,$ &
$\,\hat{S}^{2}\,(0.22)\,$ & $\,\hat{S}^{2}\,(0.25)\,$ \\
\, & \,  & \,& \,& \, & \,\\
\hline
\, & \, & \,& \,& \, & \, \\
0.024 & 0.037 & 0.061 & 0.051 & 0.035 & 0.016 \\
\, &  \, & \,& \,& \, & \, \\
\hline
\end{tabular}
\caption{\it The soft gap survival probability factor, $\hat S^2$,
for the Two Pomeron Fusion amplitude, corrected by the hard rescattering, 
for different values of $\,\alpha_s\,$ at $\sqrt{s}=14\,\,TeV\,$.}
\label{Ampl8}
\end{center}
\end{table}

Numerical values of the survival probability, based 
on \eq{sp}, are given in Tab.~\ref{Ampl8} for
different values of $\alpha_s$.  The variation of $\hat S^2$ with
$\alpha_s$ is a result of the modification, due to 
the correction term, of the shape in impact parameter~$b$ of the 
hard amplitude.

The definition of the survival probability, \eq{sp}, which we used
is different from the definition of $\hat S^2$ in  
\cite{KMR1}--\cite{KMR7}, \cite{KMR11,KMR12}.
In the case when we consider only the Two Pomeron Fusion term 
and the Gaussian form of the form factors, the definitions of 
$\hat S^2$ in \eq{sp} and in Ref.~\cite{KMR11} lead to the same 
numerical result. In the case of a general shape of the form-factors, 
however, we find it necessary to use the general formula for the soft
gap survival factor of \eq{bcorr}, \eq{final} and~\eq{sp}.

\section{Discussion and future strategies}

The hard rescattering correction to the Two Pomeron Fusion amplitude, 
even after including unitarization of the single BFKL exchange, was found 
to be large. In particular, for a more 
realistic choice of the strong coupling constant, $\alpha_s =0.25$, in the
triple Pomeron vertex, the correction exceeds significantly the Two Pomeron
Fusion term. As a consequence, we have to conclude that we have found 
another potential source of a significant suppression of the exclusive Higgs 
boson production at the LHC. 

As we have said before, we cannot consider our results as representing 
a reliable numerical final answer. In our calculation we have 
considered only a  sub-class of the screening corrections which -- 
in a BFKL-based description of diffractive Higgs production -- should 
be included, and in evaluating these diagrams, we have made severe 
approximations.   
In this paper, we have restricted ourselves to Pomeron fan diagram 
configurations in the exchanged screening system which 
lead to the Balitsky-Kovchegov evolution of the gluon density.
Such a choice is a natural starting point for an estimate of the 
amplitude beyond the Two Pomeron Fusion contribution, but it does not 
exhaust the full correction, and, in future steps, more complex topologies of 
the interacting Pomerons have to be considered.
Moreover, we have approximated the BK evolution kernel to be local 
in the impact parameter plane. 
In momentum space, this amounts to restricting all Pomerons to carry zero 
momentum transfer: such an approximation 
can be justified for the scattering on a sufficiently large nucleus; in the 
case of the proton, however, it looks somewhat questionable.  
Finally, we neglected the contribution from low mass excitations 
of the {\em target} proton in the intermediate state, which somewhat 
reduces the estimated magnitude of the rescattering correction.

Returning to the selection of diagrams, in our previous choice of 
contributions one recognizes a strong asymmetry between {\em projectile} and 
{\em target} protons, and it is obvious that a whole class of contributions 
is missing. In physical terms, the Pomerons that give rise to the BK fan 
diagrams propagate
in the colour field of the {\em projectile} proton and should be screened, too 
(examples are shown in Fig.~\ref{multipom}). It is expected that these new 
corrections will continue to come with alternating signs, 
i.e. the first screening correction of the screening 
Pomeron enters the amplitude with the positive sign etc.
Subsequent iterations of this kind lead
to a whole tower of corrections. For the total cross-section of 
nucleus-nucleus collisions an effective theory of interacting 
BFKL Pomerons~\cite{MB2,MB3,MB4}, beyond the fan diagram approximation, has 
been formulated. Imposing a semiclassical approximation that
resums Pomeron diagrams exemplified in Fig.~\ref{multipom}a, 
and restricting all Pomerons to have zero momentum transfer, 
numerical solutions have been found. 

\begin{figure}[h]
\begin{minipage}{8.0 cm}
\epsfxsize=7.0 cm
\epsfysize=5.5cm
\leavevmode
{\Large\bf a)}\\[3mm]
\hbox{ \epsffile{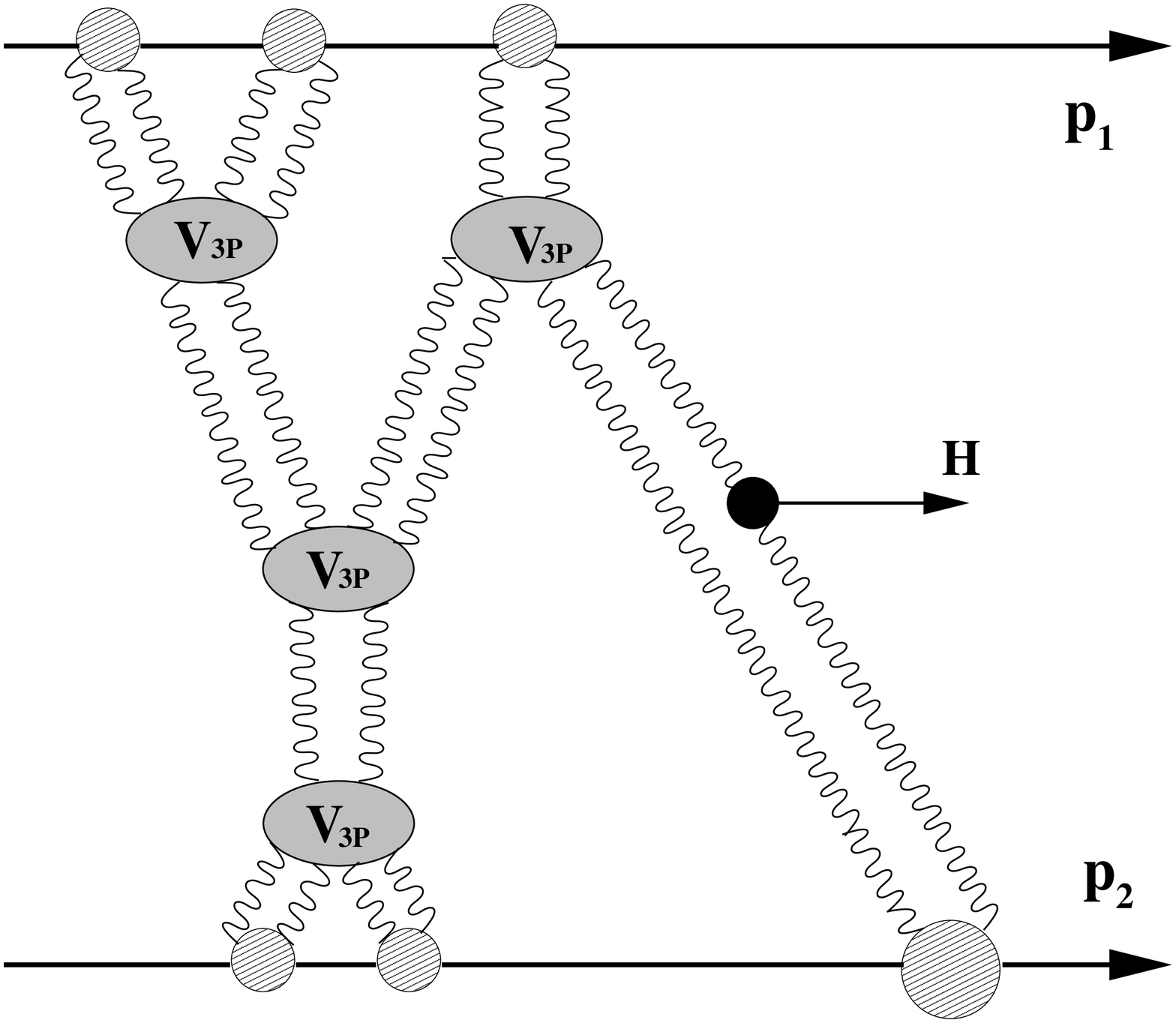}}
\end{minipage}
\hspace{0.5cm}
\begin{minipage}{8.0cm}
\leavevmode
\epsfxsize=7.0 cm
\epsfysize=5.5cm
{\Large\bf b)}\\[3mm]
\hbox{ \epsffile{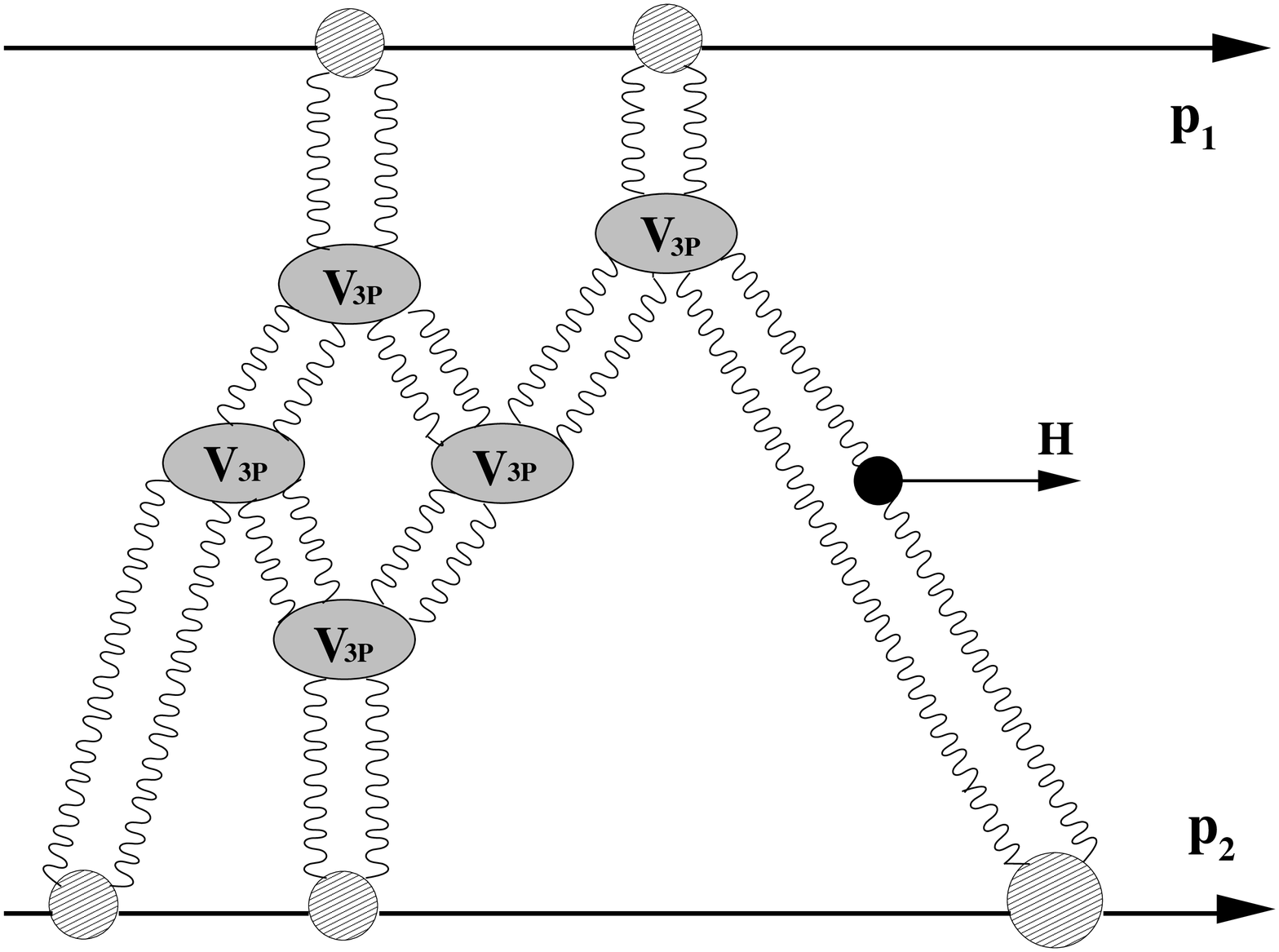}}
\end{minipage}
\caption{\it Examples of diagrams of interacting BFKL Pomeron field theory 
that contribute to the exclusive Higgs boson production: a) a diagram, without
closed Pomeron loops, belonging to the class of diagrams that were studied
in Refs.~\cite{MB2,MB3,MB4} in the case of nucleus-nucleus scattering;
b) a diagram with a closed Pomeron loop.}
\label{multipom}
\end{figure}

All these corrections can be derived from an effective 
action for interacting Pomerons of Refs.~\cite{MB2,MB3,MB4}. 
Thus, a natural continuation of the present study could be 
an analysis of the hard rescattering in this framework. 
At present an evaluation of the interacting Pomeron diagrams of the type
exemplified in Fig.~\ref{multipom}a, analogous to the treatment of 
nucleus-nucleus scattering, might be realistic, 
whereas the diagrams with closed Pomeron loops 
(see Fig.~\ref{multipom}b) are, to our understanding, still beyond reach.
Nevertheless, a careful treatment of the effective interacting BFKL Pomeron 
field theory in the semi-classical limit should be capable to provide a more 
reliable estimate of the absorption of the hard gluonic ladders than
the one performed in the present paper. 

An important point in our analysis is the role of the saturation scale.
Up to now, the analysis of diffractive Higgs production had been based upon 
a rather strict separation between hard and soft contributions: the hard 
Two Pomeron Fusion amplitude and the soft rescattering, 
encoded in the survival probability factors. In the first part of our 
hard rescattering analysis, where a single BFKL Pomeron had been considered, 
we found a strong sensitivity to the infrared scale $k_0$ which, at first 
sight, seemed to invalidate the BFKL Ansatz for the rescattering. 
However, the existence of a saturation scale, which for $x$ values being as 
small as $x \approx 10^{-6}$ lies above the infrared region, 
removes this dominance of the small-momentum region and hence supports 
our perturbative approach. In other words, for the production of a Higgs 
boson which, at the production 
vertex, provides the large momentum scale of $M_H$, the perturbative nature 
of the scattering amplitude seems not be 
limited to the production vertex but seems to `spread out' further away into 
the rescattering part of the amplitude.             
             
\section{Summary}

In this paper we have investigated hard rescattering corrections to the 
exclusive Higgs boson production at the LHC. 
In a first step, the absorptive correction was given by the 
exchange of a single BFKL ladder which couples to the hard
gluonic ladder through the triple Pomeron vertex. This amplitude has turned 
out to be strongly sensitive to the infrared region.  
In our main part, we then have 
replaced the single BFKL ladder by a Balitsky-Kovchegov gluon density.
Due to emergence of a saturation scale in the 
perturbative domain, the infrared behavior stabilizes.
We have found that the dominant contribution to the hard rescattering 
correction comes high mass diffractive excitations of the 
the {\em projectile} proton, indicating a clear separation of this 
contribution from that of low mass intermediate states of the standard soft 
rescattering.

The correction was found to be negative, with the absolute value 
being close or even exceeding the leading Two Pomeron Fusion contribution. 
This implies that the effect is significant. However, a more precise 
estimate of the cross sections is not possible until yet more screening 
corrections are evaluated and certain approximation can be 
lifted. Nevertheless, our results seem to question the simple 
picture of the $\;pp \;\to\; p\,H\,p\;$ process, being described by two hard 
gluonic ladders that fuse into the Higgs boson and leaving space only 
to a soft rescattering of the protons.
We have also discussed possibilities of improving, within the framework of 
the interacting BFKL Pomeron field theory, 
the description of the exclusive Higgs production cross section.

In order to arrive at a numerical estimate of the exclusive Higgs boson 
production cross section at the LHC, we have calculated the soft gap 
survival probability for the scattering amplitude containing both the 
Two Pomeron Fusion term and the rescattering correction.
For this we have used the dependence of the amplitude on the 
transverse position; due to the different shapes of the Two Pomeron Fusion 
term and of the correction, the gap survival factors were found to depend on 
the strength of the triple Pomeron coupling.

In our final numerical estimate of the exclusive Higgs boson 
production cross section we found that the cross section may be 
smaller, and its theoretical uncertainty may be larger than 
previously expected.  

The method developed in this paper may be also applied to
other exclusive diffractive production processes in $pp$ or $p\bar p$ 
collisions at the Tevatron or at the LHC. 
As examples, let us mention the exclusive production of 
di-jets or the exclusive production of heavy scalar mesons in the central 
region. We also view the results of this paper as a step 
towards developing a better understanding 
of the effective field theory of interacting BFKL Pomerons and of its
applications to high energy $pp$ and $p\bar p$ collisions.

\section*{Acknowledgments}
We thank Markus Diehl, Jeff Forshaw, Valery Khoze and Misha Ryskin for 
useful discussions. S.B.\ thanks the Minerva foundation for a fellowship,
K.K.\ is supported by the Graduiertenkolleg Zuk\"{u}nftige Entwicklungen 
in der Teilchenphysik. L.M.\ gratefully acknowledges the support of the grant 
of the Polish State Committee for Scientific Research No.\ 1~P03B~028~28.

\end{document}